\documentclass[aps,prb,twocolumn,showpacs,groupedaddress]{revtex4}
\usepackage{graphicx}% Include figure files

\begin{document}

\preprint{APS/123-QED}

\title{Resistivity studies under hydrostatic pressure on a low-resistance
	variant of the quasi-2D organic superconductor
	$\kappa$-(BEDT-TTF)$_{2}$Cu[N(CN)$_{2}$]Br: \\
	quest for intrinsic scattering contributions}

\author{Ch. Strack}
\author{C. Akinci}
\author{B. Wolf}
\author{M. Lang}
\affiliation{Physikalisches Institut, J.W. Goethe-Universit\"{a}t Frankfurt(M), FOR 412, D-60054 Frankfurt am Main, Germany}

\author{J.A. Schlueter}
\affiliation{Materials Science Division, Argonne NL, Argonne, Illinois 60439, USA}

\author{J. Wosnitza}
\affiliation{Institut f\"{u}r Festk\"{o}rperphysik, TU Dresden, 01099 Dresden, Germany}

\author{D. Schweitzer}
\affiliation{3. Physikalisches Institut, Universit\"{a}t Stuttgart, D-70550 Stuttgart, Germany}

\author{J. M\"{u}ller}
\affiliation{Center for Materials Research and Technology, Florida State University, Tallahassee, Florida 32306-4351, USA}

\date{\today}

\begin{abstract}
	Resistivity measurements have been performed on a low (LR)-
	and high (HR)-resistance variant of the
	$\kappa$-(BEDT-TTF)$_{2}$Cu[N(CN)$_{2}$]Br
	superconductor. While the HR sample was synthesized following
	the standard procedure, the LR crystal is a result of a
	somewhat modified synthesis route. According to their residual
	resistivities and residual resistivity ratios, the LR crystal
	is of distinctly superior quality. He-gas pressure was used to
	study the effect of hydrostatic pressure on the different
	transport regimes for both variants. The main results of these
	comparative investigations are (i) a significant part of the
	inelastic-scattering contribution, which causes the anomalous
	$\rho(T)$ maximum in standard HR crystals around 90 K, is
	sample dependent, i.e. extrinsic in nature, (ii) the abrupt
	change in $\rho(T)$ at T$^{*} \approx$ 40 K from a strongly
	temperature-dependent behavior at T $>$ T$^{*}$ to an only
	weakly T-dependent $\rho(T)$ at T $<$ T$^{*}$ is unaffected by
	this scattering contribution and thus marks an independent
	property, most likely a second-order phase transition, (iii)
	both variants reveal a $\rho$(T) $\propto$ AT$^{2}$ dependence
	at low temperatures, i.e. for T$_{c} \leq$ T $\leq$ T$_{0}$,
	although with strongly sample-dependent coefficients A and
	upper bounds for the T$^{2}$ behavior measured by T$_{0}$. The
	latter result is inconsistent with the T$^{2}$ dependence
	originating from coherent Fermi-liquid excitations.
\end{abstract}


\maketitle

\section{Introduction}
	Organic charge-transfer-salts, based on the electron\-donor
	molecule BEDT-TTF
	(bis\-ethylene\-dithio\-tetra\-thiafulvalene) --- or simply ET ---
	form layered structures consist\-ing of al\-ter\-nat\-ing
	sheets of conducting (ET)$_{2}^{+}$ cations and insulating
	anions X$^{-}$. Within this class of materials, the
	$\kappa$-phase (ET)$_{2}$X salts with X = Cu[N(CN)$_{2}$]Cl,
	Cu[N(CN)$_{2}$]Br and Cu(NCS)$_{2}$ are of particular interest
	due to the variety of electronic phases encountered as a
	function of hydrostatic pressure or anion
	substitution. According to the conceptual phase diagram
	proposed by Kanoda, the ground state of the system is
	controlled by the parameter W/U$_{eff}$, i.e., the width of
	the conduction band W relative to the effective on-site
	Coulomb repulsion U$_{eff}$, a ratio which can be changed by
	hydrostatic pressure or chemical substitutions
	\cite{Kan:97a}. This conceptual phase diagram implies that the
	antiferromagnetic insulator X = Cu[N(CN)$_{2}$]Cl and the
	correlated metal X = Cu[N(CN)$_{2}$]Br lie on opposite sites
	of a bandwidth-controlled Mott-transition. The region across
	this metal-to-insulator transition has been explored in great
	detail by employing pressure studies of various magnetic
	\cite{lef:00}, transport \cite{ito:96, lim:03} and acoustic
	\cite{four:03} properties. These studies confirm earlier
	results \cite{schir:91, sush:91} which revealed that at a
	pressure of 300-400 bar, i.e., above the critical region of
	coexistence of insulating and metallic phases \cite{lef:00,
	lim:03}, the X = Cu[N(CN)$_{2}$]Cl salt shows the same highly
	unusual resistivity profile $\rho$(T) as the Cu[N(CN)$_{2}$]Br
	system at ambient pressure \cite{kin:90}. Three distinct
	transport regimes have been identified \cite{lim:03}: (i) a
	\textit{semiconducting} high-temperature range, (ii) a
	\textit{bad metal} behavior at intermediate temperatures with
	a strongly temperature-dependent $\rho$(T) and a pronounced
	maximum around 90 K that marks the crossover to regime (i),
	and (iii) a $\rho \propto$ AT$^{2}$ behavior at low
	temperatures preceding the superconducting transition at
	T$_{c}$. Various explanations have been proposed for the
	different transport regimes. Suggestions for the anomalous
	resistance maximum include an order-disorder-transition of the
	ethylene endgroups of the ET molecules \cite{par:89, kun:93,
	tan:99} and a crossover from localized small-polaron to
	coherent large polaron behavior \cite{wan:00}, see also
	\cite{lan:04} for earlier arguments on the resistance
	anomaly. Alternatively, the \textit{"bad metal"} regime (ii)
	together with the T$^{2}$ dependence at low temperatures have
	been linked to the strongly correlated nature of the electrons
	\cite{mer:00, lim:03}. Within a dynamical mean-field (DMFT)
	approach, Merino et al. \cite{mer:00} predicted a smooth
	crossover from coherent Fermi-liquid excitations with $\rho
	\propto$ T$^{2}$ at low temperatures to incoherent
	(\textit{bad metal}) excitations at higher temperatures. Using
	such DMFT calculations for a simple Hubbard model, Limelette
	et al. \cite{lim:03} recently attempted to provide even a
	quantitative account for the $\rho$(T) behavior of pressurized
	X = Cu[N(CN)$_{2}$]Cl over an extended temperature range
	covering almost all three of the above-cited transport regimes
	(i)-(iii).

	It is fair to say, however, that despite the intensive efforts
	from both experimental and theoretical sides to explain the
	anomalous state above T$_{c}$, its nature still remains
	puzzling. In that respect, a deeper understanding of the
	unusual $\rho$(T) behavior would be of paramount importance
	given that the inelastic-scattering mechanism, which causes
	the electrical resistivity of a superconductor above T$_{c}$,
	is usually identical to the relevant pairing interaction.

	In this paper, we report resistivity measurements on a low
	(LR)- and standard high(HR)-resistance variant of the X =
	Cu[N(CN)$_{2}$]Br super\-conductor. These comparative studies,
	which include measurements under hydrostatic pressure,
	disclose striking differences between both variants. Our
	results demonstrate that --- in contrast to conventional metals
	obeying Matthiessen's rule --- extrinsic factors such as
	disorder, defects or impurities may strongly affect the
	inelastic scattering contribution in the present molecular
	conductors.

\section{Experimental}
	The temperature dependence of the resist\-ivity was measured
	by employing a standard four-terminal ac-technique operating at a
	frequency of 17 Hz. A maximum current of 10~$\mu$A was used to
	avoid self-heating. The electrical contacts to the crystal
	were made by 25~$\mu$m Cu wires attached to the sample by
	graphite paste. Typical contact resistances were $\leq$
	10~$\Omega$. Owing to the large in-plane vs. out-of-plane
	resistivity anisotropy in these materials and the irregular
	shape of the crystals, an accurate determination of the
	in-plane resistivity $\rho_{\parallel}$ is very difficult, see
	e.g. \cite{bur:94, har:97, sin:02}. As pointed out in these
	references, those "in-plane" data derived from a standard
	measurement geometry with four contacts on the same face of
	the crystal, almost always contain a significant interlayer
	component $\rho_{\perp}$. Thus, most reliable resistivity
	data, free of such mixing effects, can be obtained from
	out-of-plane measurements. To rule out errors which might
	originate in an inhomogeneous current flow in our
	four-terminal out-of-plane measurements, comparative
	investigations using a six-terminal configuration were
	conducted and found to deviate by not more than 4 $\%$ at
	maximum. For the latter measurement geometry, the current had
	been fed through the crystal by two pairs of terminals (the
	outer two of three terminals) attached to opposite crystal
	surfaces assuring these surfaces to be equi-potential
	planes. These $\rho_{\perp}$ data enable even a quantitative
	comparison with corresponding results on other crystals to be
	performed. For the in-plane measurements, $\rho_{\parallel}$,
	the current contacts were placed on opposite end surfaces of
	the crystal. A He-gas-pressure technique was used to ensure
	hydrostatic pressure conditions. The measurements were
	performed at a low sweep rate of 0.1 K/min to guarantee
	thermal equilibrium.

	The HR single crystal of
	$\kappa$\-(BEDT-TTF)$_{2}$Cu[N(CN)$_{2}$]Br was synthesized at
	the Argonne Nat\-ional Laboratory following the standard
	procedure as described elsewhere \cite{kin:88}. The LR crystal
	was obtained by solving 60 mg BEDT-TTF, 80 mg
	tetra\-phenyl\-phosphonium\-dicyanamid (Ph$_{4}$PN(CN)$_{2}$)
	and 20 mg CuBr in a mixture of 80 ml tetra\-hydrofuran (THF)
	and 20 ml ethylen\-glycol (EG). The solution was filled in a
	three-chamber electrolyte cell. The crystals were then grown
	at a current of 35~$\mu$A and a voltage of 1.3 V applied over
	a period of 14 days.

\section{Results}
	Figure 1 shows resistivity profiles $\rho$(T) for the LR and
	HR $\kappa$-(ET)$_{2}$Cu[N(CN)$_{2}$]Br single crystals at
	various pressures. The data have been normalized to their
	room-temperature values taken at ambient pressure. The figure
	discloses striking differences in the charge transport for
	both variants: Instead of the semiconducting increase in
	$\rho_{\perp}$ at higher temperatures and the pronounced
	maximum around 90 K that characterizes standard samples and is
	also present in the HR crystal studied here (Fig. 1c), the LR
	crystal (Fig. 1a) remains metallic below 300 K. Rather its
	resistivity profile reveals a weak reduction upon cooling down
	to about 120 K, below which it starts to drop more
	rapidly. This shoulder near 100 K is likely to be a remnant of
	the resistivity hump in the HR sample. In fact, as can be seen
	in Fig. 1b, a maximum around 100 K shows up in the in-plane
	resistivity data for the LR crystal as well. For the
	resistivity anisotropy, $\rho _{\perp}/\rho_{\parallel}$, our
	measurements reveal a lower limit of about 100 at room
	temperature.

\begin{figure}[htb]
\includegraphics*[width=0.70\columnwidth]{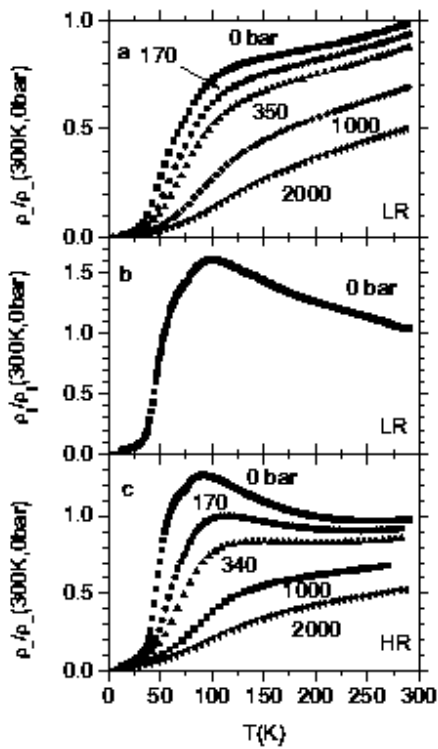}
\caption{
	Temperature dependence of the resistivity of single
	crystalline $\kappa$-(BEDT-TTF)$_{2}$Cu[N(CN)$_{2}$]Br at
	various hydrostatic pressure values up to 2000
	bar. Measurements were performed on the low-resistance (LR)
	crystal with current perpendicular (a) and parallel (b) to the
	highly conducting planes and for a standard high-resistance
	(HR) sample perpendicular to the planes (c).
}
\label{fig:1}
\end{figure}

	Apart from these sample-dependent contributions, the $\rho$(T)
	data for both crystals exhibit a sharp dip at T$_{g}$ = 77
	K. This anomaly has been assigned to a glass transition
	associated with a freezing of orientational degrees of freedom
	of the ethylene endgroups \cite{su:98b, mue:02}. With
	increasing pressure, the out-of-plane resistivity for both
	crystals becomes substantially reduced. This effect is most
	strongly pronounced at intermediate temperatures 40 K
	$\lesssim$ T $\lesssim$ 200 K, with a relative reduction
	$\rho^{-1}\Delta\rho/\Delta p = \rho(T, p_{0}=0)^{-1}[\rho(T,
	p_{0}) - \rho(T, p)]/(p_{0} - p)$ for p = 170 bar
	corresponding to about -(360 $~\pm$ 20) $\%$/kbar at 50 K and
	-(180 $\pm$ 15) $\%$/kbar at 80 K for the HR crystal. A
	somewhat smaller, though still very large, pressure response
	of\\ -(250 $\pm$ 20) $\%$/kbar (50 K) and -(120 $\pm$ 10)
	$\%$/kbar (80 K) is found for the LR crystal. At higher
	temperatures, i.e., T = 200 K and 250 K, the effect of
	pressure becomes substantially reduced reaching values of -(45
	$\pm$ 5)$\% $/kbar and -(35 $\pm$ 5) $\%$/kbar, respectively,
	which is about the same for both crystals.

	Figure 2 shows the low-temperature out-of-plane resistivity
	data for both crystals on expanded scales. For the LR crystal
	(Fig. 2a), the midpoint (50 $\%$ point) of the resistivity
	drop at ambient pressure is at 12.2 K with a 10--90 $\%$
	width of only 0.2 K. With increasing pressure, the transition
	shifts to lower temperatures and broadens progressively. Using
	the midpoint as a measure of T$_{c}$, we find an initial
	pressure coefficient of dT$_{c}$/dp$\mid_{p \rightarrow 0}$ =
	-(2.6 $\pm$ 0.2) K/kbar. These numbers have to be compared
	with T$_{c}$ = 12.0 K, a 10--90 $\%$ width of 0.4 K and
	dT$_{c}$/dp$\mid_{p \rightarrow 0}$ = -(2.4 $\pm$ 0.2) K/kbar
	for the HR crystal (Fig.  2b). The pressure coefficient of
	T$_{c}$ for both crystals is in excellent agreement with the
	results of previous pressure studies yielding pressure
	coefficients of -2.4 K/kbar \cite{schir:91} and -2.8 K/kbar
	\cite{sush:91}.

\begin{figure}[htb]
\label{fig:2}
\includegraphics*[width=0.70\columnwidth]{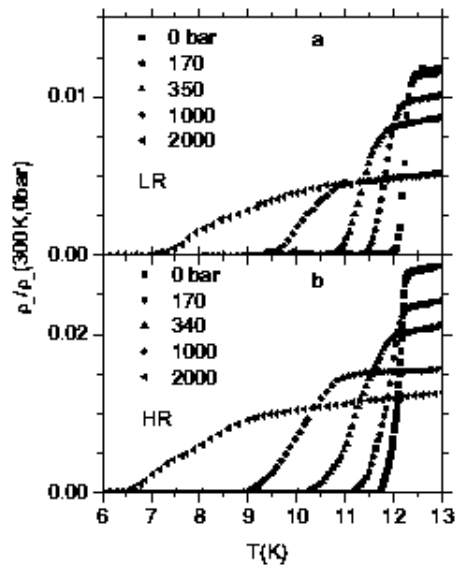}
\caption{
	Low temperature out-of-plane resistivity data at various
	hydrostatic pressure values up to 2000 bar for the low
	(LR)(a)- and high (HR)(b)- resistance variant of
	$\kappa$-(BEDT-TTF)$_{2}$Cu[N(CN)$_{2}$]Br.
}
\end{figure}

	Common to the data sets for the LR and HR crystals in Fig. 1
	is the almost abrupt change in $\rho$(T) from a strongly
	temperature-dependent behavior at intermediate temperatures to
	an only weakly temperature dependent $\rho$(T) at low
	temperatures.

\begin{figure}[htb]
\label{fig:3}
\includegraphics*[width=0.70\columnwidth]{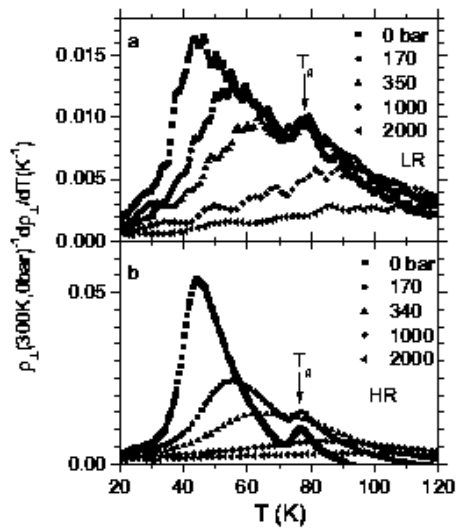}
\caption{
	Temperature derivative of the out-of-plane resistivity data
	for the low (LR) (a)- and high (HR)(b)- resistance variants of
	$\kappa$-(BEDT-TTF)$_{2}$Cu[N(CN)$_{2}$]Br at various
	pressures. Arrows indicate the glass-transition temperature at
	T$_{g}$ associated with frozen-in disorder of the ethylene
	endgroups.
}
\end{figure}

	This becomes even more clear in Fig. 3 where the derivative
	d$\rho_{\perp}$/dT is plotted for the LR (Fig. 3a) and HR
	(Fig. 3b) crystals at different pressure values. For both
	samples, we find a pronounced maximum in d$\rho$/dT at about
	the same temperature T$_{max}$ = 44 K in accordance with
	previous results on an HR crystal \cite{bur:92}. With
	increasing pressure, the maximum becomes reduced in size,
	rounded and shifted to higher temperatures. At a pressure of p
	= 2 kbar, the maximum has been suppressed almost
	completely. The sharp peak at the high-temperature side of the
	d$\rho$/dT maximum in Fig. 3 reflects the glass
	transition. Its position is almost identical for both crystals
	with T$_{g}$ = 77 K. With increasing pressure, the signature
	of the glass-transition becomes weaker while its position
	remains almost unaffected up to p = 170 bar and, for the LR
	crystal, even up to p = 350 bar. The data yield an upper limit
	for the pressure coefficient of T$_{g}$ of dT$_{g}$/dp$\mid_{p
	\rightarrow 0} \geq$ -0.6 K/kbar. At higher pressures p $\geq$
	1 kbar, however, no indication of the glass transition can be
	resolved any more.

\begin{figure}[htb]
\label{fig:4}
\includegraphics*[width=0.70\columnwidth]{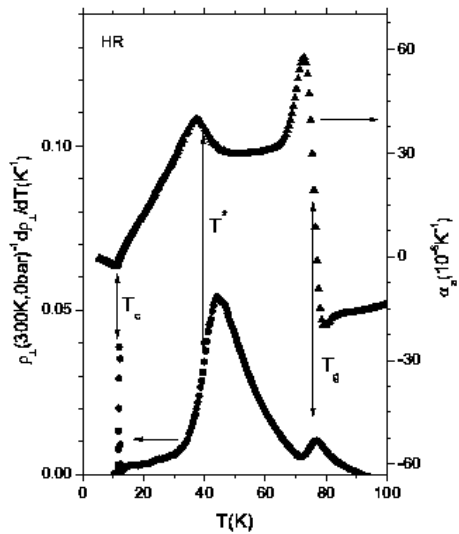}
\caption{
	Temperature derivative of out-of-plane resistivity data,
	d$\rho _{\perp}$/dT, (left scale) and thermal expansion
	results (right scale) taken from reference ~\cite{mue:02} for
	two different high-resistance
	$\kappa$-(BEDT-TTF)$_{2}$Cu[N(CN)$_{2}$]Br crystals plotted on
	the same temperature scale. The resistivity data for T $\leq$
	10.5 K have been omitted for clarity. Arrows indicate
	positions of the superconducting (T$_{c}$) and glass
	transition (T$_{g}$), as well as for the anomaly at T$^{*}$.
}
\end{figure}

	In Fig. 4 we compare the temperature dependence of the
	d$\rho_{\perp}$/dT data of Fig. 3b with those of the
	coefficient of thermal expansion measured along the in-plane
	a-axis, $\alpha_{a}$, on a similar HR crystal
	\cite{mue:02}. The figure discloses a clear correspondence of
	the features in d$\rho$/dT with the phase-transition-like
	anomalies observed in $\alpha_{a}$(T) at T$_{c}$ = 12 K,
	T$_{g}$ = 77 K and T$^{*} \simeq$ 40 K \cite{mue:02}. More
	precisely, as indicated by the arrow at T$^{*}$, it is the
	midpoint of the low-T side of the d$\rho$/dT maximum which
	coincides with the transition temperature T$^{*}$ determined
	from $\alpha$(T).  Using the midpoint as a measure of T$^{*}$,
	the data in Fig. 3 can be used to determine the pressure
	dependence of T$^{*}$. For pressures p $\leq$ 350 bar, this
	criterion yields about the same pressure coefficient of
	dT$^{*}$/dp$\mid_{p \rightarrow 0}$ = + (35 $\pm$ 7) K/kbar
	for both variants. This value slightly exceeds the pressure
	effect of about + 25 K/kbar reported by Frikach et
	al. \cite{fri:00} who followed the position of the pronounced
	minimum in the sound velocity as a function of pressure.

\begin{figure}[htb]
\label{fig:5}
\includegraphics*[width=0.70\columnwidth]{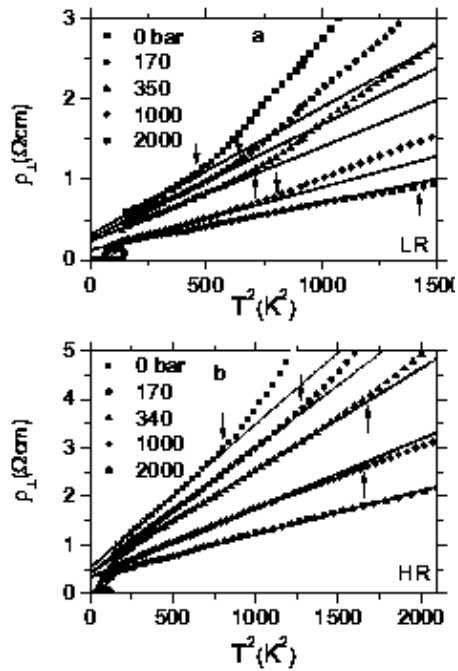}
\caption{
	Low-temperature out-of-plane resistivity data for the low
	(LR)(a)- and high (HR)(b)-resistance variant of
	$\kappa$-(BEDT-TTF)$_{2}$Cu[N(CN)$_{2}$]Br under various
	pressures plotted as a function of T$^{2}$. Arrows mark the
	temperatures where the data deviate by more than 2$\%$ from
	the straight lines.
}
\end{figure}

	Figure 5 shows the low-temperature inter-layer resistivity
	data in a $\rho$(T) vs. T$^{2}$ representation. In accordance
	with published results \cite{bur:92, dre:97, su:98a}, the
	normal-state resistivity of the HR crystal (Fig. 5b) follows a
	$\rho$(T) = $\rho_{0}$ + AT$^{2}$ behavior over an extended
	temperature range. From Fig. 5b, we derive a coefficient
	A$^{HR}$ = (3 $\pm$ 0.6) m$\Omega$cmK$^{-2}$ and a residual
	resistivity $\rho_{0}^{HR}$ = (530 $\pm$ 100) m$\Omega$cm. The
	error bars account for uncertainties implied in determining
	the geometrical factor. A T$^{2}$ dependence is also found for
	the LR crystal (Fig. 5a) although with markedly smaller values
	for the coefficient A and the residual resistivity of A$^{LR}$
	= (1.6 $\pm$ 0.4) m$\Omega$cmK$^{-2}$ and $\rho_{0}^{LR}$ =
	(320 $\pm$ 80) m$\Omega$cm, respectively. In addition, Fig 5
	discloses significantly different validity ranges for the
	T$^{2}$ law for both variants. Using a 2$\%$ deviation of the
	straight lines in Fig. 5 as a measure for the upper boundary
	T$_{0}$ of the T$^{2}$ dependence, we find ambient-pressure
	values of T$_{0}^{LR}$ = (23 $\pm$ 0.5) K and T$_{0}^{HR}$ =
	(30 $\pm$ 0.5) K for the LR and HR crystal, respectively. As
	indicated by the arrows in Fig. 5, both variants reveal a
	strongly non-linear, and, for the HR crystal, even a
	non-monotonous, change of T$_{0}$ with pressure. We note that
	the analysis of the in-plane data at ambient pressure of the
	LR crystal in Fig. 1b reveals a T$_{0}^{LR}$ value which is
	identical to that derived from the out-of-plane resistivity.

\begin{figure}[htb]
\label{fig:6}
\includegraphics[width=0.70\columnwidth]{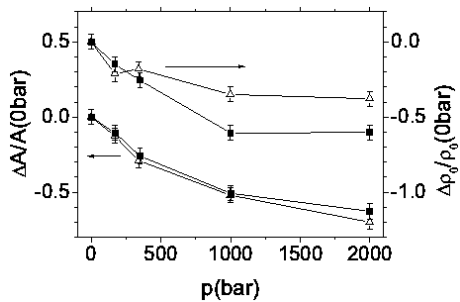}
\caption{
	Relative change of the coefficient A, $\Delta$A = A(p) -
	A(0bar), and the residual resistivity $\rho_{0}$,
	$\Delta\rho_{0} = \rho_{0}(p) - \rho_{0}$(0bar), with pressure
	for the low (LR) (filled squares)- and high-resistance (HR)
	(open triangles) variant of single crystalline
	$\kappa$-(BEDT-TTF)$_{2}$Cu[N(CN)$_{2}$]Br.
}
\end{figure}

	Figure 6 compiles the relative changes of the coefficient A,
	$\Delta$A/A(p = 0) = A(0)$^{-1}$[A(p) - A(0)] (left scale) and
	the residual resistivity $\rho_{0}$, $\Delta \rho_{0}/\rho_{0}
	= \rho_{0}^{-1}[\rho_{0}(p) - \rho_{0}(0)]$ (right scale) as a
	function of pressure. For the coefficient A, we find almost
	identical behavior for both crystals with a stronger reduction
	at small pressures and a weak pressure dependence at p $\geq$
	1 kbar. A similar tendency can be inferred also for the
	residual resistivity although here the pressure effect for the
	HR crystal is somewhat smaller and there is no significant
	pressure dependence for p $\geq$ 1 kbar.

\section{Discussion}
	As described in the experimental section, the LR and HR
	variants of $\kappa$-(ET)$_{2}$Cu[N(CN)$_{2}$]Br are the
	results of somewhat different preparation routes which may
	cause variations in the purity of the materials, i.e. the
	concentration and the nature of incorporations \cite{Cu:II},
	and the structural perfection. The latter refers to the degree
	and character of structural disorder. Although detailed
	comparative structural investigations of the LR and HR
	crystals which require highest-resolution techniques have not
	been performed yet, a general characterization of the crystals
	studied here is feasible on the basis of the present transport
	measurements.

	According to the residual resistivities, derived from an
	extrapolation of the normal-state $\rho _{\perp}$(T) to T = 0,
	of $\rho_{0}$ = (320 $\pm$ 80) m$\Omega$cm (LR) and (530 $\pm$
	100) m$\Omega$cm (HR), and the residual resistivity ratios RRR
	= $\rho_{\perp}$(300K)/$\rho_{0\perp}$ of 158 (LR) and 67
	(HR), the LR crystal is of distinctly superior quality. The HR
	crystal studied here, however, appears to be representative
	for most of the $\kappa$-(ET)$_{2}$Cu[N(CN)$_{2}$]Br crystals
	studied so far which had been prepared according to the
	standard procedure. These crystals yield room-temperature
	resistivities $\rho_{\perp}$(300K) and RRR values of 50-70
	$\Omega$cm and 50-65, respectively \cite{bur:92, su:98b}.

	The possibility of internal strain, which might account for
	the suppression of the anomalous resistivity maximum at
	intermediate temperatures for the LR crystal, can be safely
	discarded due to both the high T$_{c}$ value and the very
	narrow 10--90 $\%$ transition width of only 0.2 K. The latter
	is a factor of 2 smaller to that which is usually encountered
	for this salt \cite{bur:92, su:98b} and which is found also
	for the present HR crystal. At the same time, both variants
	behave almost identical as for the glass-transition
	temperature T$_{g}$ = 77 K, although the signature at T$_{g}$
	in the interlayer resistivity, i.e. the additional scattering
	contribution to $\rho _{\perp}$ for T $<$ T$_{g}$, is stronger
	for the HR crystal. This might indicate a reduced fraction of
	frozen-in disordered ethylene groups in the LR compared to the
	HR sample.

	The most obvious difference between the HR and LR crystals
	highlighted in figure 1 is the distinct reduction of the
	$\rho$(T) maximum at intermediate temperatures. Yet a remnant
	of this feature, though much less strongly pronounced, is
	still present for the LR sample, where it gives rise to an
	unusual $\rho$(T) anisotropy with a metallic-type resistivity
	in the out-of-plane component but a semiconducting-like
	behavior for the in-plane resistivity. We note, however, that
	the resistivity anisotropy $\rho _{\perp}/\rho_{\parallel}$ of
	about 100 at room temperature, derived from the present
	experiments (cf. Fig. 1) as compared to an anisotropy ratio in
	excess of 1000 reported by Buravov et al. \cite {bur:92}, for
	an HR crystal employing an improved Montgomery method,
	indicates that the present $\rho_{\parallel}$ data determined
	by using a standard four-terminal measurement geometry still
	contain a significant interlayer component $\rho_{\perp}$.

	A strongly reduced, though finite, scattering contribution
	around 90 K in the LR crystal is in line with the observation
	of a significant reduction of the still extraordinarily strong
	pressure response of the resistivity at intermediate
	temperatures compared to that of the HR crystal. In fact,
	sample-to-sample variations in this scattering contribution,
	though substantially reduced in size, may also be found for
	crystals prepared along the same alternative route that led to
	the present LR crystal. Here we mention the sample studied in
	reference ~\cite{Wos:99} and two further crystals of this salt
	explored in the course of the present investigation. These
	samples revealed a clear correlation between the residual
	resistivity ratio and the size of the resistivity around 90 K:
	upon increasing the RRR ratio from 84 over 89 to 158,
	$\rho_{\perp}$(90 K) continuously decreases. For the crystal
	studied in reference \cite{Wos:99} yielding RRR = 193,
	$\rho_{\perp}$(T) is almost identical to that found for the
	present LR crystal.

	The above observation that the anomalous scattering
	contribution centered around 90 K differs strongly depending
	on the preparation conditions and becoming reduced in size
	upon increasing the sample quality, indicates that it is
	extrinsic in nature.  Moreover, these results demonstrate that
	disorder or defects may induce drastic changes in the
	temperature-dependent part of the resistivity, i.e. the
	inelastic-scattering contributions. Such a behavior is highly
	unusual and at variance with what is known from ordinary
	metals, where the scattering due to disorder or impurities
	manifests itself in an increase of the residual resistivity
	only. This raises the fundamental question on how and to what
	extent disorder- or defect-induced potentials may affect the
	inelastic scattering of $\pi$ electrons in the present
	molecular conductors.

	Apart from these differences related to the anomalous
	resistivity contribution around 90 K, both variants behave
	identically as to the drastic change in their resistivity at T
	= T$^{*} \simeq$ 40 K from a range characterized by a strongly
	T-dependent $\rho$(T) at T $>$ T$^{*}$ into a low-temperature
	regime, where $\rho$(T) varies only weakly with
	temperature. It was found that the anomaly in d$\rho$/dT
	coincides with the phase-transition-like feature observed in
	the coefficient of thermal expansion $\alpha$(T). Such a
	direct correspondence of anomalies in transport and
	thermodynamic quantities is not expected for a crossover
	behavior between two different regimes, which usually involves
	a scaling factor to map the characteristic temperatures
	T$_{\rho}$ and T$_{\alpha}$.  As an example, we mention the
	signatures of the Kondo effect in $\rho$(T) and $\alpha$(T) in
	heavy fermion compounds, see, e.g. \cite{steg:90}. Rather the
	coincidence of distinct anomalies in d$\rho$/dT and
	$\alpha$(T) indicates that this feature marks a cooperative
	phenomenon.

	Anomalous behavior around T$^{*}$ has been also identified in
	various thermal \cite{sush:91, mue:02, bur:92, mur:90,
	sas:02}, magnetic \cite{kat:92, may:94, kaw:95, wzi:96},
	elastic \cite{fri:00}, and optical properties
	\cite{sas:04}. Various explanations have been proposed as to
	the nature of the T$^{*}$ anomaly, including the formation of
	a pseudo-gap in the density of states \cite{kat:92, may:94,
	kaw:95, DeS:95}, a crossover from a coherent Fermi liquid at
	low temperatures into a regime with incoherent excitations at
	high temperatures \cite{mer:00, lim:03}, a density-wave-type
	instability \cite{mue:02, sas:02, lan:03}, as well as an
	incipient divergence of the electronic compressibility caused
	by the proximity to a Mott transition \cite{four:03}. The
	present resistivity results, which for the HR crystal confirm
	published data \cite{bur:92}, clearly demonstrate that the
	position of the T$^{*}$ anomaly is unaffected by the strength
	of the additional scattering contribution giving rise to the
	resistivity hump at intermediate temperatures, and thus marks
	an independent feature. In addition, the sharpness of the
	anomaly in d$\rho$/dT and its direct mapping with the jump in
	the coefficient of thermal expansion makes it very unlikely
	that T$^{*}$ merely reflects a crossover between two different
	transport regimes---an assumption which underlies some of the
	above theoretical models. Rather it indicates that T$^{*}$
	reflects a phase transition into a symmetry broken
	low-temperature state.

	Turning now to the $\rho$ = $\rho_{0}$ + AT$^{2}$ behavior for
	T $\leq T_{0} < $T$^{*}$, our study reveals a relative change
	with pressure of the coefficient A which is quite similar for
	the LR and HR crystals. This indicates that it is the same
	scattering mechanism which governs the low-temperature
	$\rho$(T) behavior for both systems. However, the size of A is
	substantially reduced for the LR crystal reflecting a
	weakening of this scattering contribution for the
	higher-quality LR crystal.

	There has been a long-standing debate on the nature of the
	T$^{2}$ behavior in the resistivity of molecular
	conductors. In fact, a $\rho \propto$ T$^{2}$ dependence over
	an extended temperature range is not a peculiarity of the
	$\kappa$-phase (ET)$_{2}$X salts alone. It has been observed
	also for various other materials such as the
	(TMTSF)$_{2}$PF$_{6}$ and the $\beta$-(ET)$_{2}$X salts, see,
	e.g. \cite{bul:88, weg:88, weg:93}.

	The explanations proposed for the T$^{2}$ behavior in these
	materials include electron-phonon \cite{weg:94} as well as
	electron-electron interactions of the strongly correlated
	$\pi$-electron system \cite{bul:88, mer:00, lim:03}. In fact,
	such a T$^{2}$ dependence at low temperatures is
	characteristic of metals in which the dominant scattering
	mechanism is provided by the electron-electron
	interactions. Since there the coefficient A $\propto
	(m^{*})^{2} \propto (T_{F}^{*})^{-2}$, with m$^{*}$ the
	effective carrier mass and T$_{F}^{*}$ the effective Fermi
	temperature, the coefficient A scales with the square of the
	Sommerfeld coefficient $\gamma \propto m^{*} \propto
	(T_{F}^{*})^{-1}$ of the electronic specific heat C$_{el}$ =
	$\gamma$T. Such an A/$\gamma^{2}$ = const. behavior within a
	given material class has been verified for different systems
	including heavy-fermion compounds and transition metals
	\cite{kad:86, miy:89}.

	The above scaling implies that upon variation of a control
	parameter x of the system, such as chemical composition or
	external pressure, the product A(x)$\cdot(T_{F}^{*}(x))^{2}$
	should stay constant. By identifying the temperature T$_{0}$,
	i.e., the upper limit of the T$^{2}$ range in the resistivity,
	with the effective Fermi energy T$_{F}^{*}$, Limelette et
	al. have verified this invariance for pressurized
	$\kappa$-(ET)$_{2}$Cu[N(CN)$_{2}$]Cl \cite{lim:03}.

	The results of the present studies, however, render such an
	interpretation unlikely. Given that the T$^{2}$ dependence is
	of electronic origin, i.e. A $\propto (T_{F}^{*})^{-2}$ and
	T$_{0} \simeq T_{F}^{*}$, the A coefficient for the LR
	variant, which is reduced by a factor of about 1.9 compared to
	that of the HR crystal, would then indicate an effective Fermi
	temperature T$_{F}^{*}$ which is larger by a factor of
	(1.9)$^{1/2} \approx$ 1.4. This is in contrast to the
	experimental observation yielding a T$_{0}^{LR}$ which is even
	reduced by a factor of about 1.3 compared to that for the HR
	crystal. Rather, our experimental finding that both T$_{0}$
	and A are strongly sample dependent while the other
	characteristic temperatures associated with the electronic
	properties such as T$^{*}$ and T$_{c}$ are not, indicate that
	the nature of the T$^{2}$ dependence is different from
	coherent Fermi-liquid excitations \cite{beta:structure}.

\section{Summary}
	Resistivity measurements under hydrostatic pressure on a
	low-resistance variant of the organic superconductor
	$\kappa$-(BEDT-TTF)$_{2}$Cu[N(CN)$_{2}$]Br have been performed
	and compared to the results on a standard high-resistance
	crystal.  The lower residual resistivity $\rho_{0}$ and the
	higher residual resistivity ratio $\rho(300K)/\rho_{0}$ for
	the low-resistance crystal clearly indicate its superior
	quality. These measurements reveal that a significant part of
	the scattering contribution which gives rise to the anomalous
	resistivity maximum around 90 K in standard high-resistance
	materials is extrinsic in nature.  Apart from this
	sample-dependent scattering contribution, however, both
	variants behave identically as to the abrupt change in
	$\rho$(T) at T$^{*} \simeq$ 40 K. The coincidence of this
	temperature with the phase-transition anomaly in the
	coefficient of thermal expansion makes it unlikely that
	T$^{*}$ marks a crossover between two different transport
	regimes but rather indicates a second-order phase
	transition. For temperatures T$\leq T_{0} < $T$^{*}$ the data
	for both crystals were found to follow a $\rho$(T) $\propto$
	AT$^{2}$. Most importantly, however, our analysis reveals
	strikingly different coefficients A and ranges of validity
	measured by T$_{0}$ for both variants. In view of the fact
	that other characteristic temperatures associated with the
	$\pi$-electron system such as T$_{c}$ and T$^{*}$ are sample
	independent, this strong variation in A and T$_{0}$ indicates
	an origin for the T$^{2}$ dependence different from coherent
	Fermi-liquid excitations. The present results demonstrate that
	for these molecular materials, sample dependent,
	i.e. extrinsic factors such as disorder or defect
	concentration, does not only change the elastic scattering
	contribution measured by the residual resistivity. Rather, the
	defect potentials may also strongly affect the
	temperature-dependent part of the resistivity, i.e. the
	inelastic scattering, indicating that Matthiessen's rule is no
	longer applicable to these materials. Consequently, the charge
	transport for available sample materials might considerably be
	affected by such extrinsic contributions. Detailed structural
	investigations on high- and low-resistance material are in
	progress to hopefully identify the nature of the above
	scattering centers. This will help to control better the
	synthesis conditions and to eventually provide materials of
	sufficiently high quality which make it possible to access the
	intrinsic transport properties of these materials.

\begin{acknowledgments}
	The authors thank M. Kartsovnik and N. Toyota for fruitful
	discussions and suggestions.  The work was supported by the
	Deutsche Forschungsgemeinschaft under the auspices of the
	Forschergruppe 412.
\end{acknowledgments}


\begin{thebibliography}{Buravov 271}
\bibitem{Cu:II}In~reference~\cite{mont:99} the influence of magnetic 
	Cu(II) incorporations on the anomalous resistivity profiles
	has been investigated but no clear correlation has been
	obtained so far.
\end{thebibliography}
\end{document}